\documentclass[12pt,a4paper]{article}
\usepackage{graphicx}
\begin{document}
\title{Selfish vs.\ Unselfish Optimization of Network Creation}
\author{Johannes J Schneider$^1$ and Scott Kirkpatrick$^2$ \\
{\small $^1$ Institute of Physics, Johannes Gutenberg University of Mainz} \\
{\small Staudinger Weg 7, 55099 Mainz, Germany} \\
{\small schneidj@uni-mainz.de} \\
{\small $^2$ School of Engineering and Computer Science} \\
{\small The Hebrew University of Jerusalem} \\
{\small Givat Ram, Jerusalem 91904, Israel} \\
{\small kirk@cs.huji.ac.il}}

\maketitle

\begin{abstract}
We investigate several variants of a network creation model:
a group of agents builds up a network between them while trying
to keep the costs of this network small. The cost function
consists of two addends, namely (i) a constant amount for each edge
an agent buys and (ii) the minimum number of hops it takes
sending messages to other agents. Despite the simplicity of
this model, various complex network structures emerge depending
on the weight between the two addends of the cost function
and on the selfish or unselfish behaviour of the agents.
\end{abstract}

{\bf PACS:} {89.75.Hc,89.75.Fb,89.20.Hh,89.65.-s,05.10.Ln}

\noindent{\it Keywords\/}:
 optimization, selfish optimization, network, internet,
 multi agent system

\section{Introduction}
The Internet, which has changed computing by making access
to information and to computing resources available to the
world, is a complex system worthy of study in its own right.
It came into being through the federation of many different networks
originally designed to serve different purposes for different communities,
and does not have strong central management.
Attempts to understand its behaviour at the lowest level,
that of connectivity and topography, have shown that the simplest model,
a graph in which sites are connected at random,
must be discarded in favor of a graph in which
the distribution of the number of links is very heavy-tailed.
Most sites have very few connections, but there is a power-law
tail containing very few, very highly-connected
sites \cite{Vespignani}.  The earliest maps of the Internet \cite{BTLwork}
tended to show a fairly tree-like structure, but these were gathered
by searching out from a single point
along shortest paths, producing essentially a minimum spanning tree to the
selected destinations. More extensive searches \cite{neu1}
show that there are many crosslinks as well. The most
recent, detailed studies of the links between subnetworks, building  blocks
of the Internet, show an average connectivity of more than six links per
site in addition to power law tails in the degree distribution
\cite{DIMEShome,neu2}.

Other networks, like the World Wide Web, networks of actors with
joint movies as edges between them, and various kinds of networks
between humans, exhibit similar properties \cite{Vespignani,neu3}.
Besides these investigations of the properties of the Internet
and of related networks,
one is interested in the basic mechanisms leading to this type of
network. The growth and shaping of a network is a stochastic process
of considerable interest, but the lack of central control makes the use
of statistical mechanical models for such complex systems suspect.
In fact, such a network appears to be a "game," in which
many independent agents manage its components to suit their own needs.
Recent attempts to analyze the implications of this thought \cite{worst}
have focused on the "price of anarchy," a catchy term given to the ratio of the
cost of the worst-case selfish but stable solution in such a game to the
"social optimum" in which all players cooperate to produce the best solution
for the system as a whole.  This ratio may be a 
constant, or may increase with $N$, the number of players in the game. 

The equilibrium in which each agent has chosen a specific
configuration for the assets
that it manages and, given the configurations chosen by the other agents,
has no incentive to change, is called a pure Nash equilibrium \cite{Nash}.
Even when the social cost is carefully defined to be the sum of the individual
costs which individual players optimize, the social optimum may not be a Nash
equilibrium, if individual incentives destabilize it.
Nonetheless, the search for a pure Nash equilibrium or more 
generally when the agents' information or choices are more restricted,
a selfish optimum, is a stochastic process of great interest both
as a variation on statistical mechanics, and as a model of how large complex
systems behave in the real world.  
 
A recent model that we shall consider is "network creation,"  introduced
in \cite{Fabrikant}
and extended to models which represent actual peer-to-peer data sharing
networks formed as overlays in the actual Internet, in \cite{Chun}.
This is the first of what may be many models in which the differences
between selfishly driven optimization and global optimization can shed light
on real-world complex systems.  Determining the full range of possible
selfishly optimal behaviour in such models is at least as difficult
as combinatorial optimization, and has in fact spawned
new complexity classes in computer science \cite{pureNash}.

\section{The Network Creation Model}
In the network creation model, one agent is assigned to each node $i$
of the graph. Each edge
between a pair $(i,j)$ of nodes is owned either by the agent on
node $i$ or the agent on node $j$, but it can be used by any agent
for transferring messages.
The cost of sending messages will depend on our model of message
traffic.  For simplicity and generality, we shall assume that each agent
must send an equal number of messages (one, without loss of generality) to
all other agents.  This uniform model will never generate the power law
tails seen in the real Internet,  but it proves to exhibit surprisingly
rich  behaviour. It requires that the graph be connected. Thus each agent
must purchase sufficient edges in order to be connected to all other
agents, but as this occurs in an asynchronous parallel process, agents can
take advantage of edges purchased by others.
Buying an edge costs an amount $\alpha$.
In this model, $\alpha$ is simply a constant and does not depend
e.g.\ on the distance between the nodes $i$ and $j$ or on the
required bandwidth for the connection.
The cost of sending each message is given by the number of hops (the number
of links that it passes over) in the shortest path between the sender and
the receiver.
The costs of one hop is an arbitrary parameter,
such that it is conveniently set to 1.

If the network is already connected, each agent still has
the problem whether it wants to buy additional
edges in order to reduce the costs induced by the number of hops
or to take a larger amount of hops into account in order to reduce
the costs incurred by the edges. The decision to which it gets at
the end depends strongly on the value of $\alpha$: if $\alpha$
is smaller than $1$, then of course a complete graph with edges
between every pair of nodes is preferrable. However, if $\alpha$
is very large, then the network is surely only a tree, such that
the condition that the graph must be connected is fulfilled and no
edge more than needed is added. The most interesting question for
us is what structures the networks created by the agents have
for intermediate values of $\alpha$. These structures will not
only depend on the value of $\alpha$ but also on the decision of
the agents when to buy an additional edge. This decision depends
also on the behaviour of the agents, whether the agents are selfish,
i.e., if they only consider whether their own costs decrease, or
not, i.e., if they consider whether the sum of the costs of all agents
decreases.

Related to this behaviour, one can write down cost functions for the
single agents:
the cost function of a selfish agent on node $i$ can be written as
\begin{equation}
{\cal H}_{\mathrm{selfish}}(i)=
\sum_{j\atop{\mathrm{agents}}} \left(\alpha \times \eta_{ij} + d_{ij}
\right)
\end{equation}
with $\eta_{ij}=1$ if the agent on node $i$ has bought an edge
to node $j$ and 0 otherwise and with $d_{ij}$ being the distance
between the nodes $i$ and $j$ measured in hops
as described above. Analogously,
the cost function of an unselfish agent on node $i$ is given as
\begin{equation}
{\cal H}_{\mathrm{unselfish}}(i)=
\sum_{j\atop{\mathrm{agents}}} \alpha \times \eta_{ij} +
\sum_{k\atop{\mathrm{agents}}}
\sum_{j\atop{\mathrm{agents}}} d_{kj} .
\end{equation}
Thus, in the unselfish scenario, each agent considers only the
costs of the own links but all the costs induced by the distances.

One can also define an overall cost function for the system which
is given as
\begin{equation}
{\cal H}_{\mathrm{total}}=
\sum_{i\atop{\mathrm{agents}}}
\sum_{j\atop{\mathrm{agents}}} \left(\alpha \times \eta_{ij} +
d_{ij}\right) .
\end{equation}
This cost function is independent of the behaviour of the agents:
it is identical with the sum over all selfish
cost functions ${\cal H}_{\mathrm{selfish}}(i)$ of the single
agents. But it is also basically equal to the cost function of an unselfish
agent, as each agent can only make decisions due to the connections
they bought on their own but not on the connections bought by
other agents.

In these distances $d_{ij}$, one can also consider the constraint that
the graph has to be connected: the agent on
node $i$ adds up the number of
hops messages take to any other node $j$ to which it is directly or
indirectly via other nodes connected and stores this number
in the $d_{ij}$. For all other nodes, which it
cannot reach, it sets $d_{ij}=L$ with $L$ being a large number.
$L$ has to be larger than $\alpha$ and the maximum possible
number of hops in the system to ensure that the graph will be connected
at the end of the simulation.

\section{Simulation Details}
Simulating this model, one mostly starts with an empty graph without
edges between the nodes, because this is the natural starting point.
For comparison, additionally to this ``from scratch''-scenario
we investigate a ``from complete''-scenario in which one starts with
a complete graph in which all pairs of nodes are directly connected
via an edge which is randomly owned by one of the nodes. Whereas the
``from scratch''-scenario can be justified by the historic development,
as there were no connections at the beginning,
the ``from complete''-scenario can be pictured as if there were a
planning meeting of all agents in which they started with a complete
graph und with edges which were already assigned to them and
they determined in a random order which edges should be
removed.

Besides the creation of a starting configuration,
a simulation must specify rules for moves, i.e., a prescription how
to change the configuration. An obvious choice for a move
from the point of view of an agent
is of course a buy/sell-move: one chooses a pair $(i,j)$ of nodes
at random with $i \neq j$.
If there is no edge between them, then the agent
on node $i$ determines whether
it is preferrable for it
to buy an edge to $j$ or not. According to the behaviour
of the agent, it will make its decision.
If the situation is improved in its view by
buying an edge then the agent on node $i$
buys an edge to $j$. However, if there is
already an edge between $i$ and $j$, then the buy/sell-move asks
the owner of the edge whether it is preferrable for it to remove it.
This buy/sell-move can also be accepted if the costs for the agent stay
the same, i.e., if the move leads to an equally good configuration,
as it is anyway nowadays in the habit of managers to
always restruct their companies. If such a restruction
does not lead to any deterioration, it is to be accepted.

Besides this buy/sell-move, which is surely a natural move, one
can also implement a switch-move: here a triple $(i,j,k)$ of
nodes is randomly chosen in the way that the agent on node $i$ owns an edge
between $i$ and $j$ and that there is no edge between $i$ and $k$.
Now the agent on node $i$ is asked whether it is preferrable
to delete the edge to $j$ and to add an edge to $k$ at the same time.
Implementing this additional move might lead to even better
configurations, as the simulation cannot get stuck at configurations
anymore which could be easily improved by this switch-move but not
by the successive application of sell- and buy-moves, as a sequence
of these is only accepted if none of these leads to any deterioration,
whereas the corresponding switch-move has only to consider the remaining
part of the sum of the improvements and of the deteriorations.

After having performed many such moves, the simulation run usually
reaches a configuration which is not changed anymore. In the case of
the unselfish behaviour of the agents, in which the simulation
is performed basically according to the total cost function of the system,
the simulations often end at a local or even the global
minimum in the energy
landscape. However, not always such a local minimum is reached due
to the ownership of a link: an agent can only delete an edge if it
is owned by it. Thus, if it is rather well but not optimally connected
because an other agent bought an edge to it and if it is not
advantageous to buy a further edge as the costs for it would be
larger than the savings in the reduction of the distances,
it will have to stay with this locally not optimal situation.
The configurations in which simulation runs
end up are some types of local minima. In optimization, one
differentiates between local and global minima: global minima
have cost function values which are optimal for the problem,
there is no configuration at all with a better cost function
value. Contrarily, local minima only have a cost function
value which is better than the cost function values of all
configurations which can be reached by the application of
one move from this local minimum. In the world of Multi Agent
Systems, a Nash equilibrium corresponds to the global optimum.
We here end up at local minima in which the simulations
get stuck and which cannot be improved by the application
of any move available to an agent.
We test for reaching such a local minimum
explicitly. Before stopping the simulation we explore all possible
moves that are available to a single agent.
It could also be that
a simulation run never gets stuck in a stable configuration because the
system can jump between equally good configurations.

Thus, we cannot let a simulation run through a possibly infinite
loop. In our
implementation, 10000 steps are performed. In each step, first
100 moves are tried in a random way, i.e., the nodes which shall buy,
sell, and switch, rsp., edges and their neighbors are selected
at random. If all of these 100 moves are rejected, then all
possibilities for moves are checked deterministically but
in a random order. This shall ensure that in each step
at least one move is accepted. If during this complete search for a move
to be accepted no acceptable move is found, then the simulation
run is already stuck in a local minimum or Nash equilibrium, such
that the simulation can be stopped ahead of time. Otherwise, the
simulation proceeds with the next step.

In our simulations, we either use only the buy/sell-move (b/s)
or use both the buy/sell-move and the switch-move
in a random order (b/s+sw) with equal probability. 
Furthermore, we either start ``from scratch'' (fs) or ``from
complete'' (fc) and the agents exhibit either a selfish or
an unselfish behaviour. Thus, we have all in all eight
different scenarios. For each of these scenarios, we consider
several values for $\alpha$, namely $0.5$, $0.7$, $1$, $1.3$,
$1.5$, $1.7$, $2$, $2.3$, $2.5$, $2.7$, $3$, $3.3$, $3.5$,
$3.7$, $4$, $5$, $6$, $7$, $8$, $9$, $10$, $11$, $12$, $13$,
$14$, $15$, $16$, $17$, $18$, $19$, $20$, $30$, $40$,
$50$, $60$, $70$, $80$, $90$, $100$, $110$, $120$, $130$,
$140$, $150$, $160$, $170$, $180$, $190$, $200$, $300$,
$400$, and $500$. For each scenario and for each value
of $\alpha$, we performed
100 simulation runs over which the results are averaged.

\section{Computational Results}
\begin{figure}\centering
\parbox{\textwidth}{
\includegraphics[width=.49\textwidth]{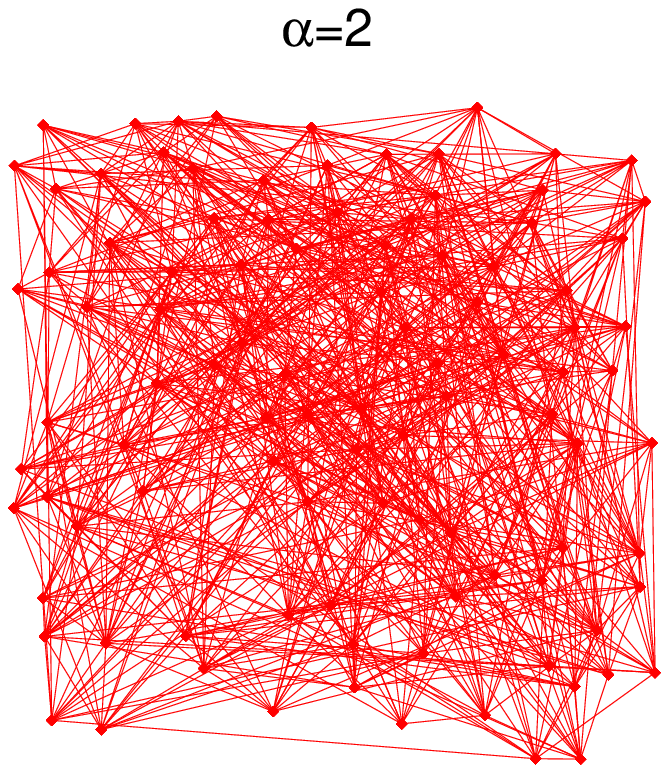}
\hfill
\includegraphics[width=.49\textwidth]{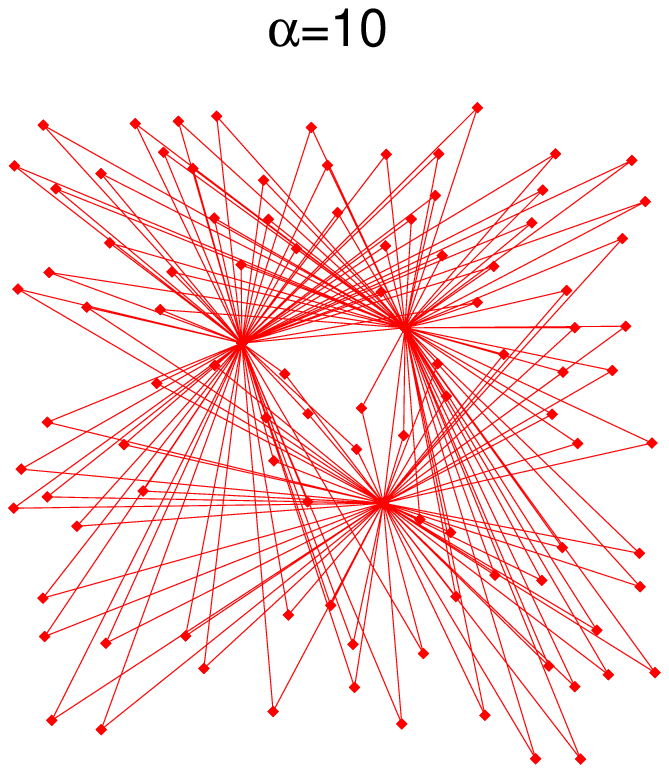}
}
\parbox{\textwidth}{
\includegraphics[width=.49\textwidth]{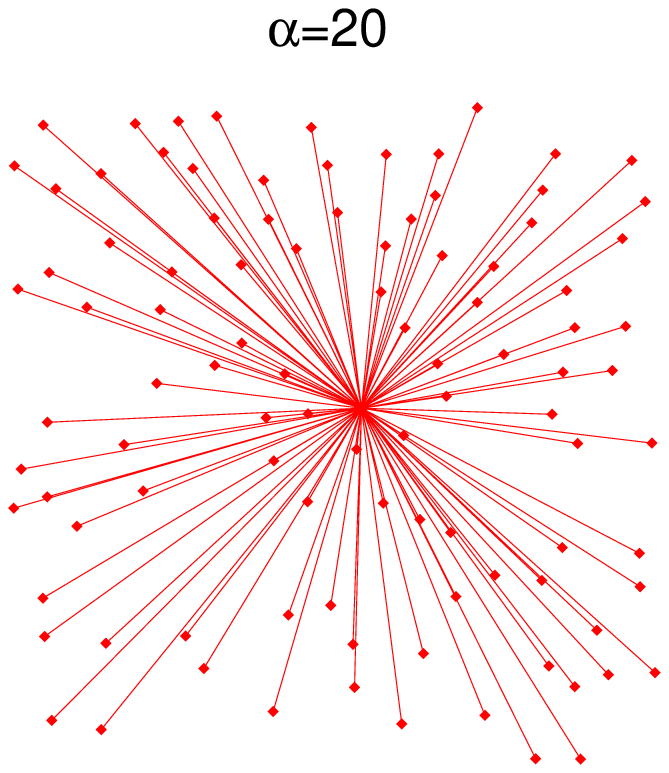}
\hfill
\includegraphics[width=.49\textwidth]{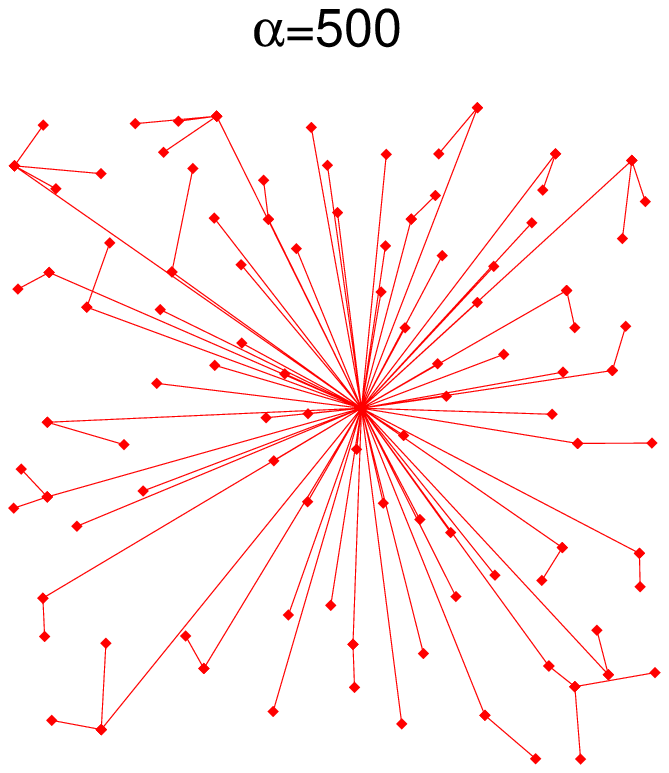}
}
\caption{Examples of resulting configurations: depending on the
value of $\alpha$, simulations
lead to different interesting final configurations.
}
\label{fig:examples}
\end{figure}

The final configurations of our simulations strongly depend
on the size of $\alpha$: obviously, the simulations end in
full connected graphs for $\alpha<1$, as buying an edge is
cheaper than letting a message perform an additional hop.
Figure \ref{fig:examples} shows typical final configurations
reached at larger values of $\alpha$. Although these examples
were found using selfish optimization, from scratch, similar results are
found starting from a complete graph, or using global optimization. For
$\alpha = 2$, the example graph found is dense, but no longer complete.
For different configurations found at $\alpha=2$, the number of
edges in the configurations varies between 99 and 917 with an average
of 613.
We often get a star with one centre node to which all other nodes are
connected for $\alpha=20$. For even larger values of $\alpha$,
the simulations mostly produce trees which, at first, are
close to stars in
structure. But also some rather unexpected structures
occur: for $\alpha=10$, we often get a structure with
three centre nodes. All other nodes are connected to two of
these three centre nodes. There are two edges between these
three nodes, such that messages need only up to two hops
to get from the sender to the receiver. In
graph theory terms, the solution for intermediate values of alpha maintains
the diameter of the network, the longest shortest path between any two
points, at two hops.

Now we want to study the behaviour of our simulations
and the final configurations statistically.

\begin{figure}\centering
\parbox{\textwidth}{
\includegraphics[width=.49\textwidth]{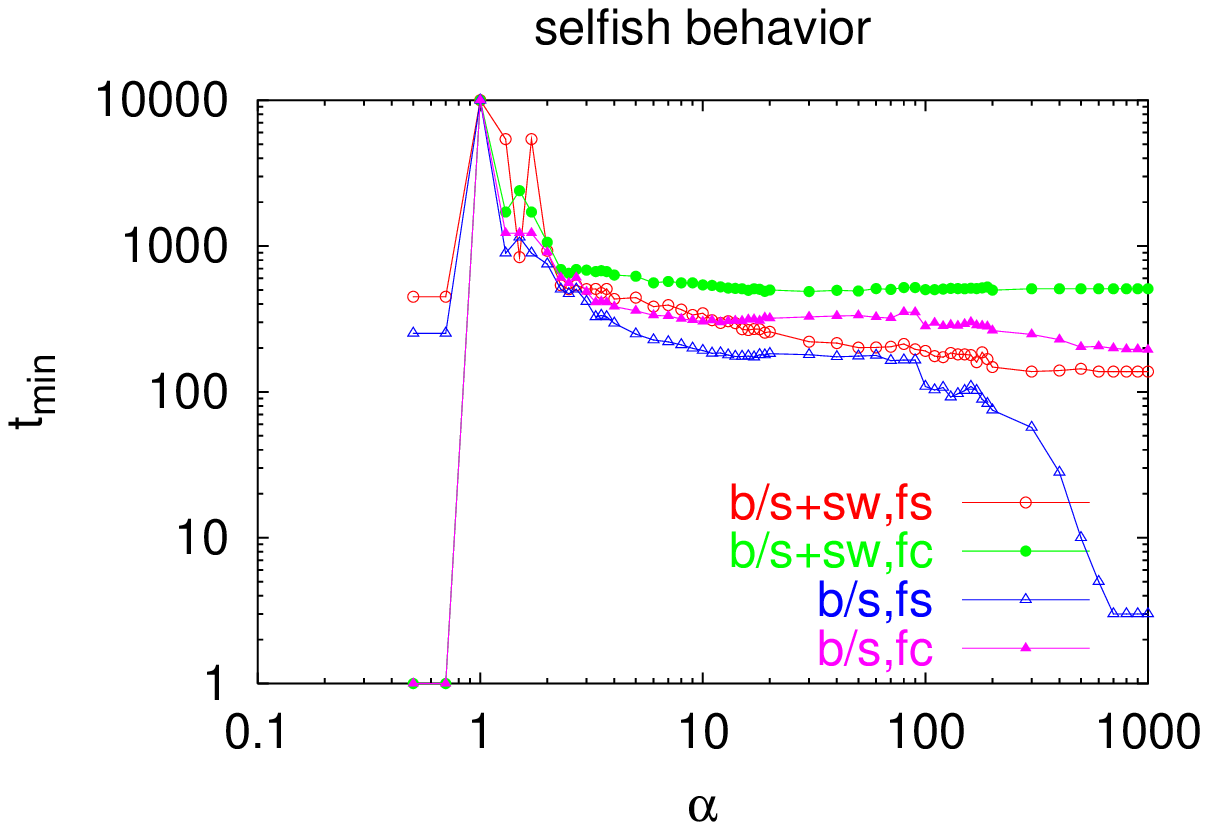}
\hfill
\includegraphics[width=.49\textwidth]{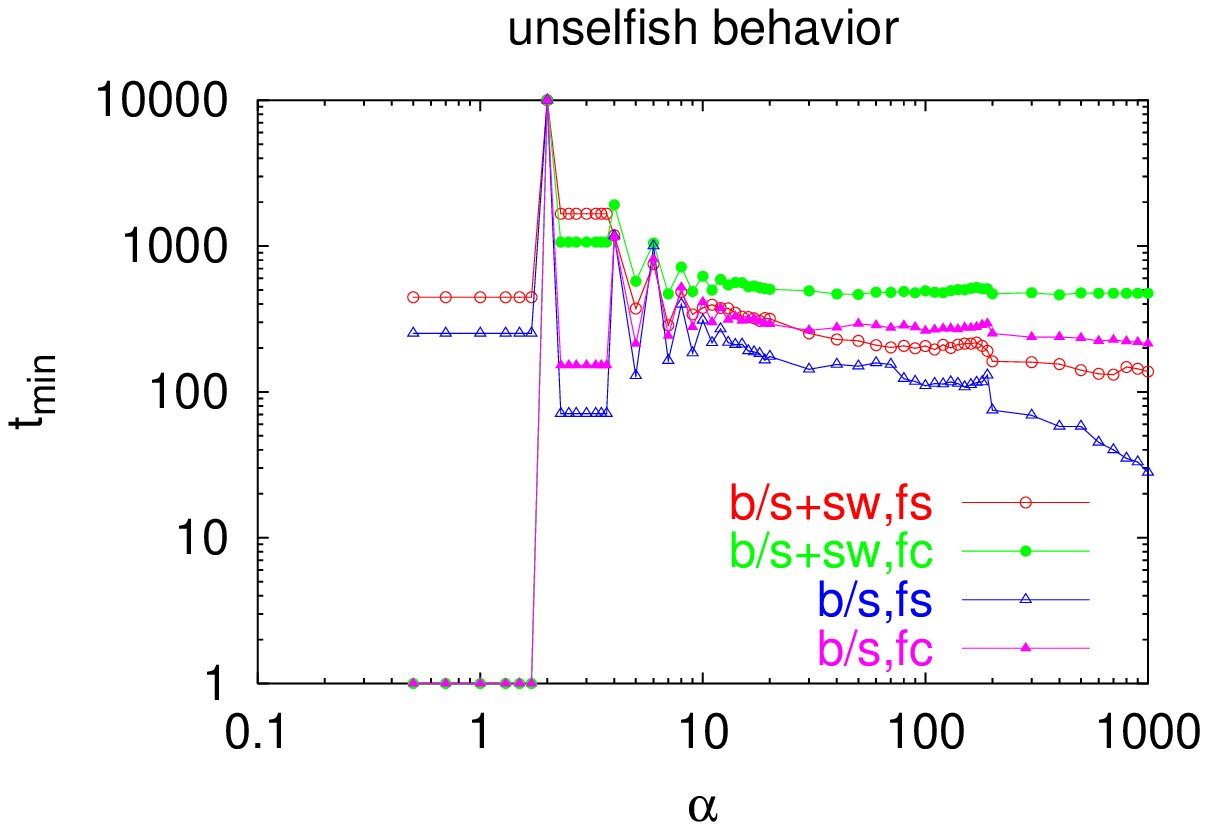}
}
\parbox{\textwidth}{
\includegraphics[width=.49\textwidth]{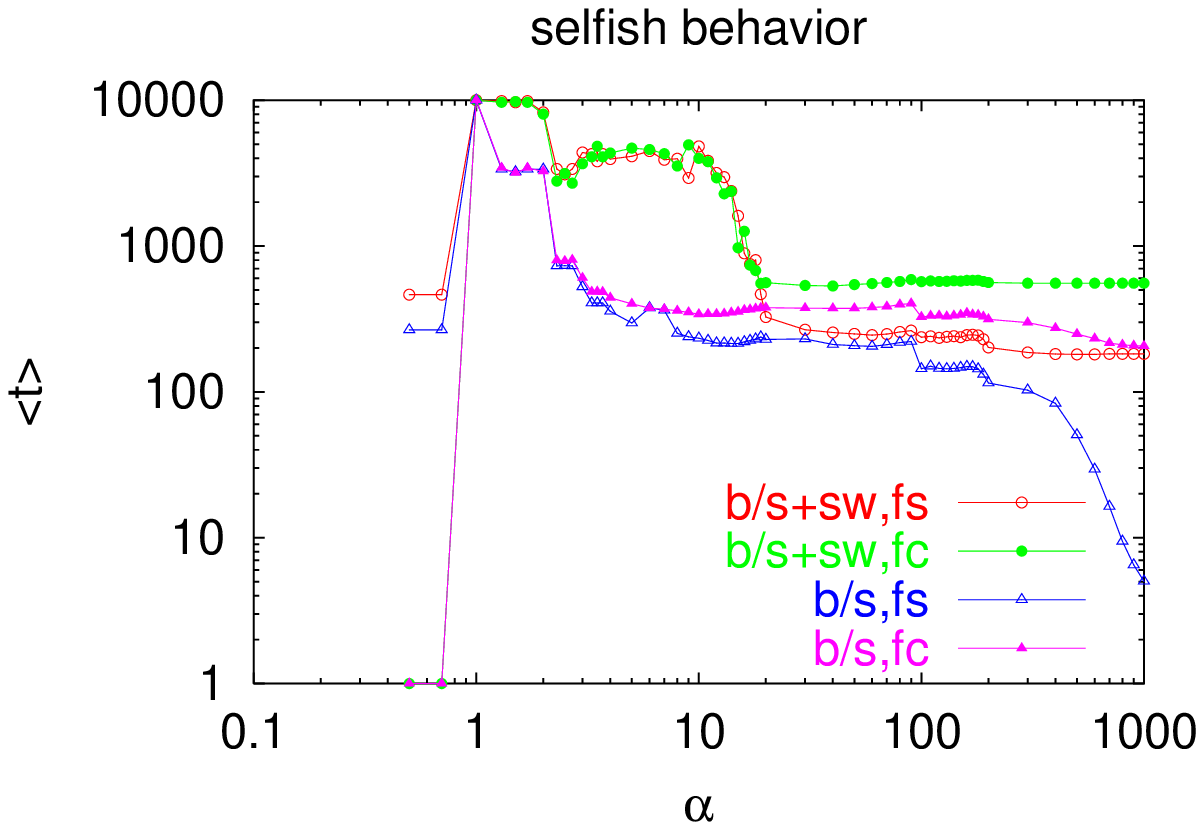}
\hfill
\includegraphics[width=.49\textwidth]{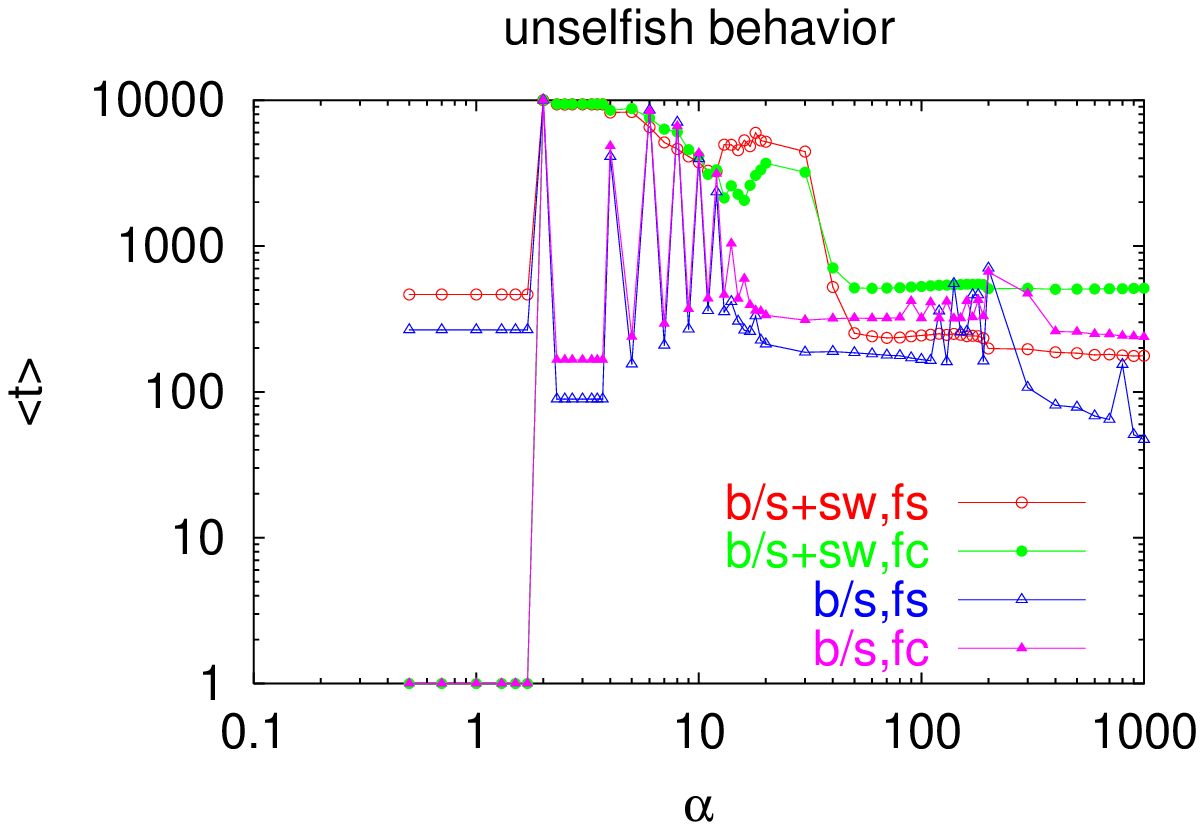}
}
\parbox{\textwidth}{
\includegraphics[width=.49\textwidth]{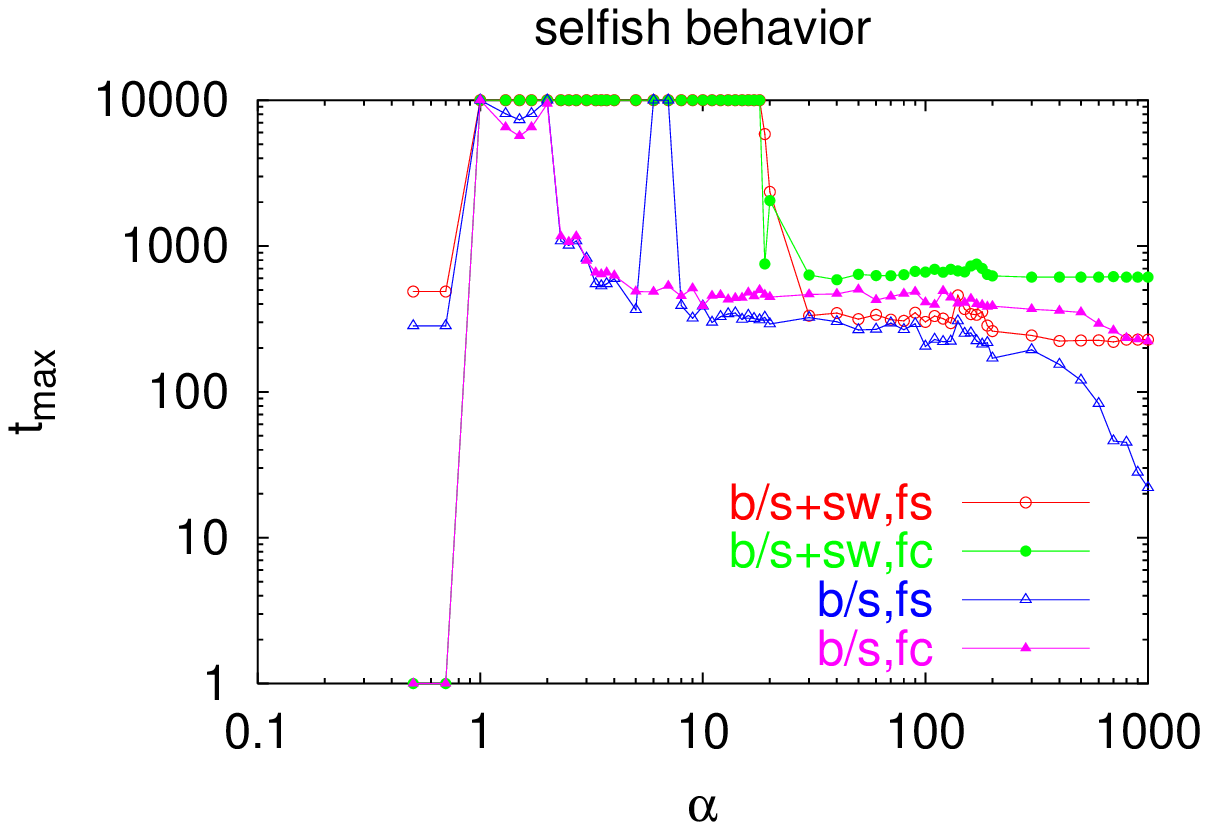}
\hfill
\includegraphics[width=.49\textwidth]{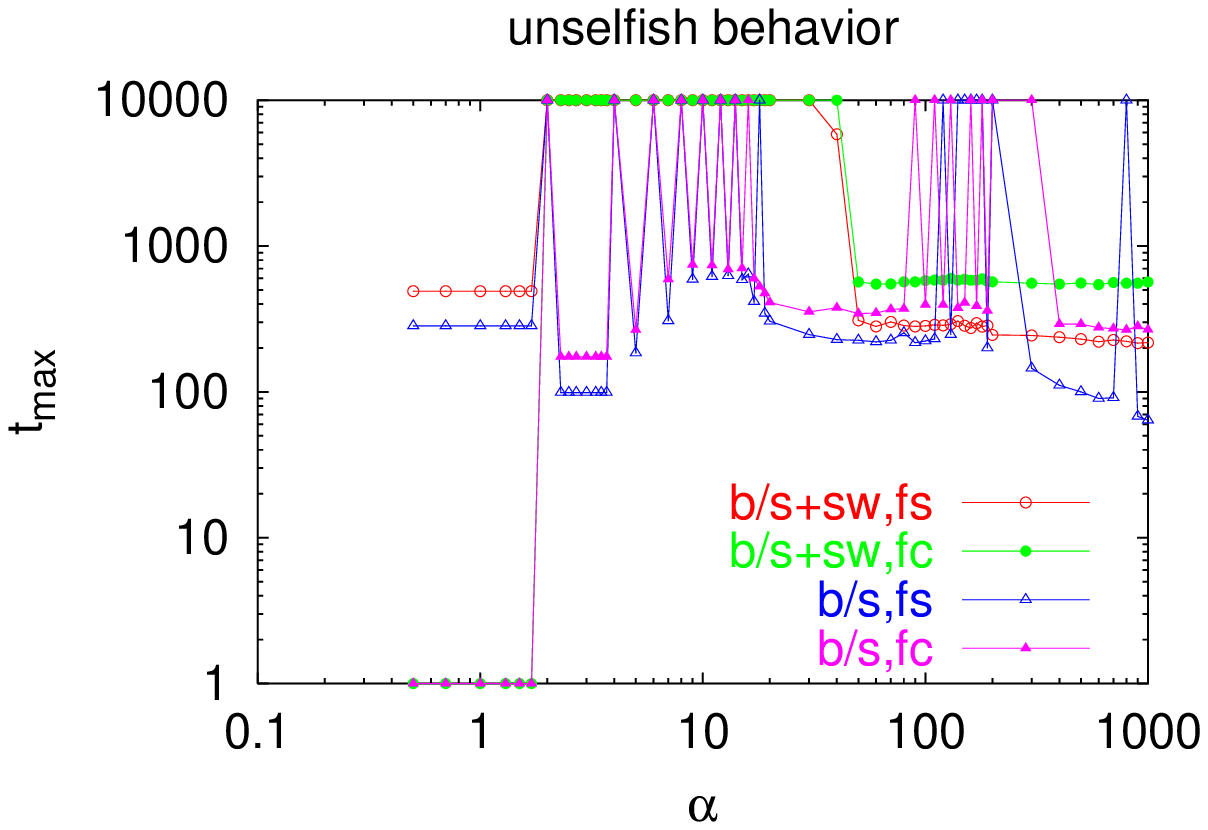}
}
\caption{Minimum (top), mean (middle), and maximum (bottom) number
of steps until the system reaches either a Nash equilibrium or
a local minimum: the results for the selfish behaviour are shown
on the left, for unselfish behaviour on the right. In each
graphics, empty symbols mark results for the ``from scratch''-scenario
(fs), filled symbols results for the ``from complete'' scenario
(fc). Results for using both moves (b/s+sw) are shown as circles,
results for using the buy/sell-move only (b/s) as triangles.
}
\label{fig:endzeit}
\end{figure}

Figure \ref{fig:endzeit} shows the minimum, mean, and maximum numbers
of steps, the simulations took. Please note that there is a maximum
of 10000 steps: if the simulation has not reached a stable Nash
equilibrium or a local minimum, it breaks after this number of steps.
It might be that such a broken simulation might have ended soon after,
but it also might have gone on forever. Furthermore
note that the minimum number of steps is always 1: thus, even if the
simulation starts in a Nash equilibrium or a local minimum, such that
no move is accepted, one step is counted. There are indeed simulations
which were started from such minima: as already mentioned, a complete
graph with edges between all pairs of nodes is optimal for $\alpha<1$.
Thus, if we start from our ``from complete''-scenario for some
$\alpha<1$, there is no move leading to any improvement, such that
the simulation can already stop at this point.
For selfish agents working on $\alpha=1$, we find that the runs
are only ended by the rule that a simulation cannot exceed 10000
steps. Here many neighboring configurations, which can be transferred
via a move to each other as the move does not lead
to a deterioration, are degenerate.

Then in the range
of intermediate values of $\alpha$, we see several structures,
of which the most interesting is the one for unselfish agents
using the buy/sell-move only: here the number of steps the simulation
needs to come to an end strongly depends on whether $\alpha$,
when it takes an integer value, is even or odd. The calculation
time is much larger for even integer values, as can be seen
both at the minimum and mean and maximum number of steps in the
range $2 \le \alpha \le 18$.
The time is often only limited by the maximum number of allowed steps,
whereas simulations with odd integer values always come to an
end in much less than 10000 steps. The reason for this
on first sight strange behaviour is that every distance $d_{ij}$
is counted twice in the Hamiltonian for the unselfish agents,
namely once as distance $d_{ij}$ from $i$ to $j$ and once as
distance $d_{ji}$ from $j$ to $i$. Now if an unselfish node
decides to add a further edge for a cost value of $\alpha$
which reduces the distances of e.g.\ $\frac{\alpha}{2}$ pairs of
nodes by an amount of 1, this move leads to an equally good
configuration. Thus, both this move and its inverse move are
accepted, leading to degenerate configurations between which
the simulation can jump endlessly. The final configurations
of these simulation runs contain several centre nodes, all other
nodes are connected to two of these centre nodes, which are
partially connected with each other.

Such degeneracies do not occur for odd values of $\alpha$,
as here adding or deleting an edge leads to either a better
or a worse configuration. If we had used a factor of $\frac12$
with which we had multiplied the sum over the distances in the
cost function of unselfish agents, we would see degeneracies
at all integer values of $\alpha$ and much quicker convergence
on the non-integral numbers in between.

Sometimes we also get an increase in the calculation time
at $\alpha\approx N$ in the unselfish case using only the
buy/sell-move. Here some of
our simulations also run into configurations which are
degenerate with their neighbors.

Please note that these times were so far given in steps.
However, there is a large difference in the computing times
between simulations with selfish agents and simulations
with unselfish agents. Working with selfish agents,
only the distances from this agent to all other agents
have to be determined. Here the calculation time for
a move goes with the order of ${\cal O}(N)$. For unselfish
agents, the new distances between all pairs of agents
have to be determined, such that there the calculation
time goes with ${\cal O}(N^2)$. Thus, we had to spend
weeks of calculation time on dozens of work stations
for our simulations of unselfish agents whereas we could
perform the simulations of selfish agents within a
comparatively short time.

\begin{figure}\centering
\parbox{\textwidth}{
\includegraphics[width=.49\textwidth]{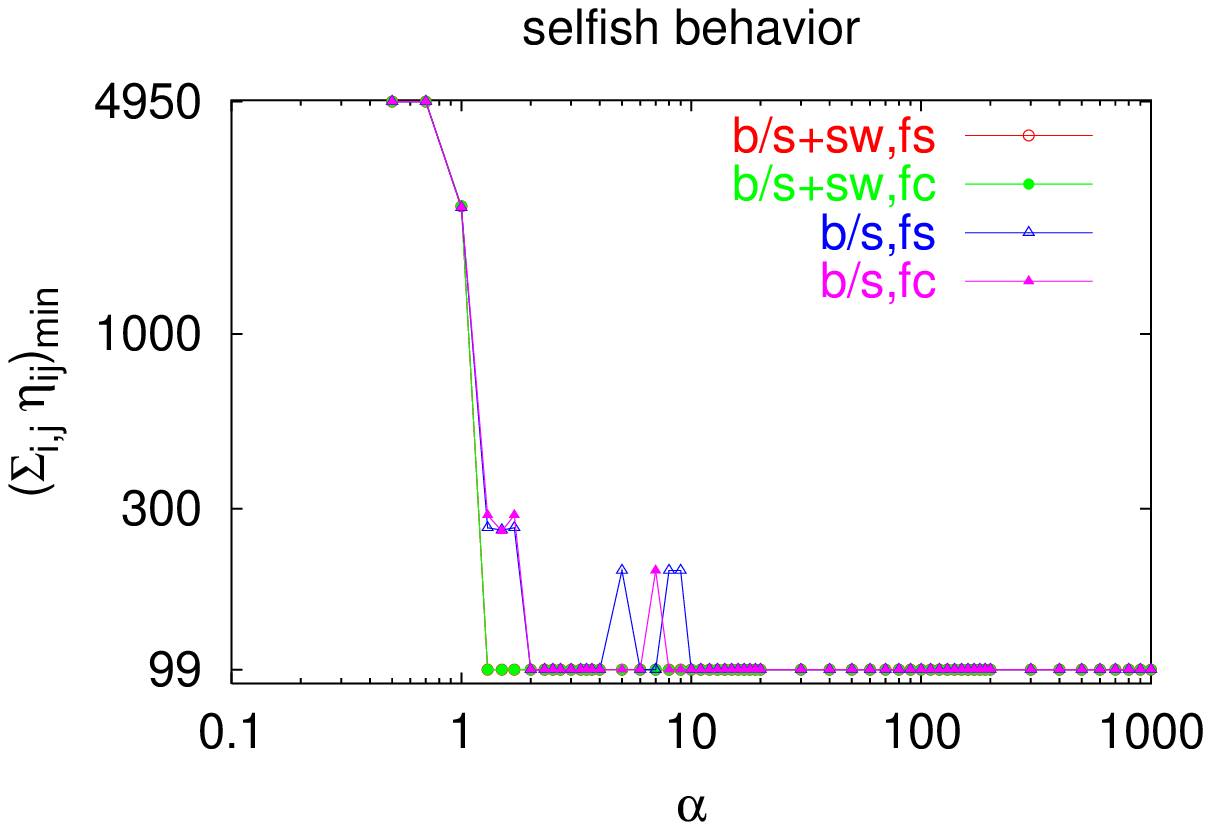}
\hfill
\includegraphics[width=.49\textwidth]{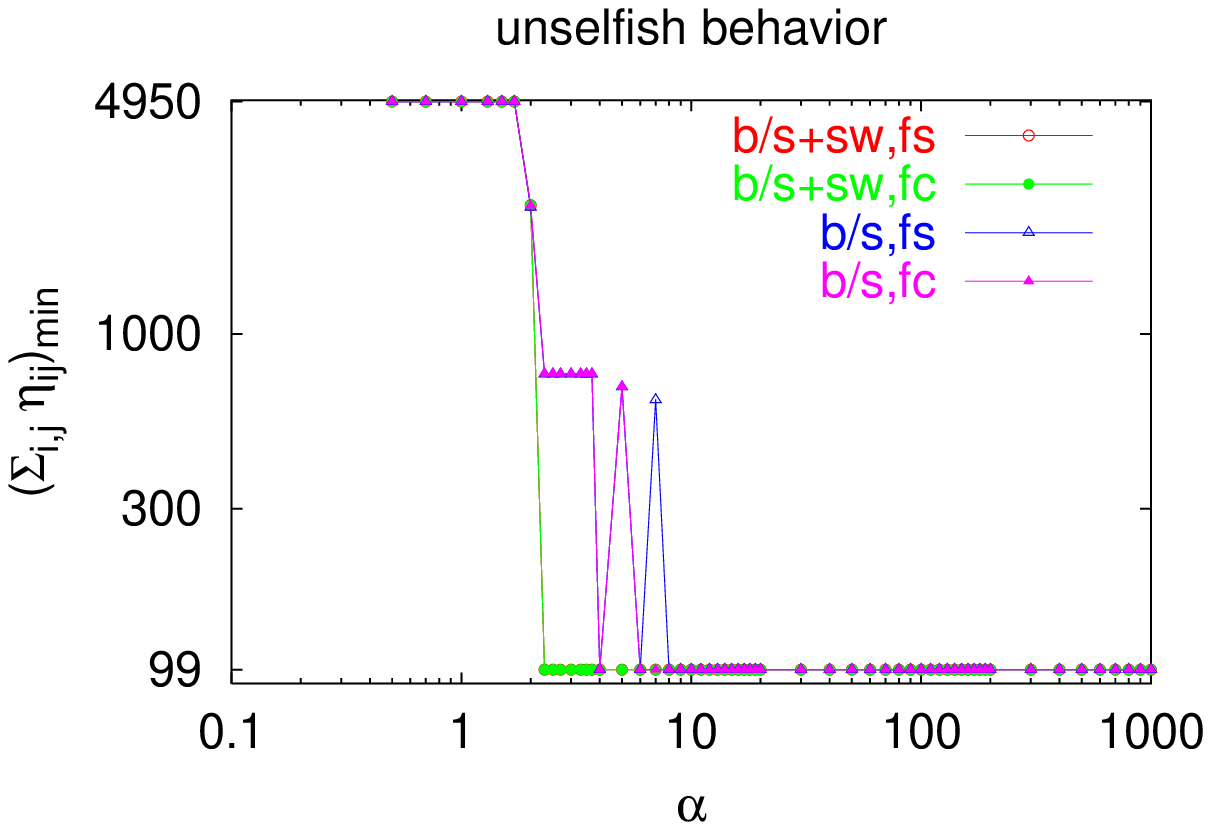}
}
\parbox{\textwidth}{
\includegraphics[width=.49\textwidth]{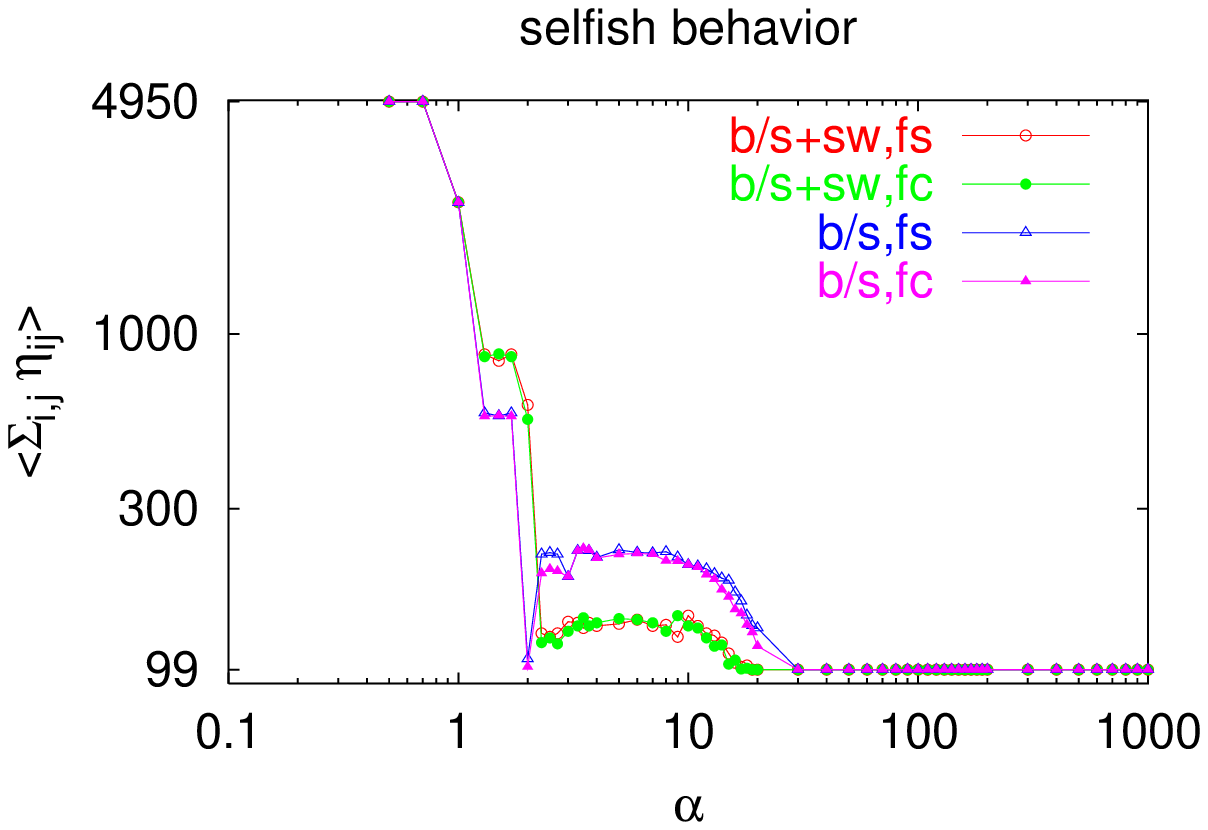}
\hfill
\includegraphics[width=.49\textwidth]{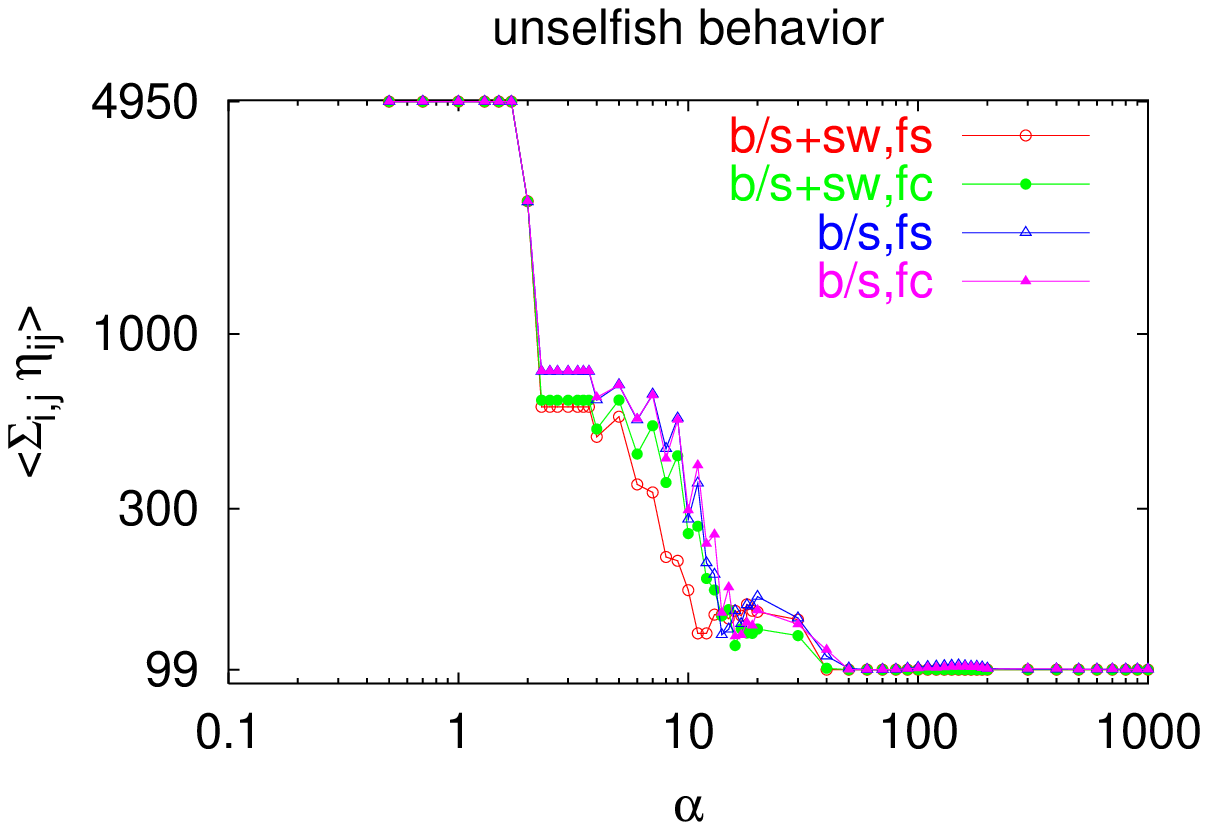}
}
\parbox{\textwidth}{
\includegraphics[width=.49\textwidth]{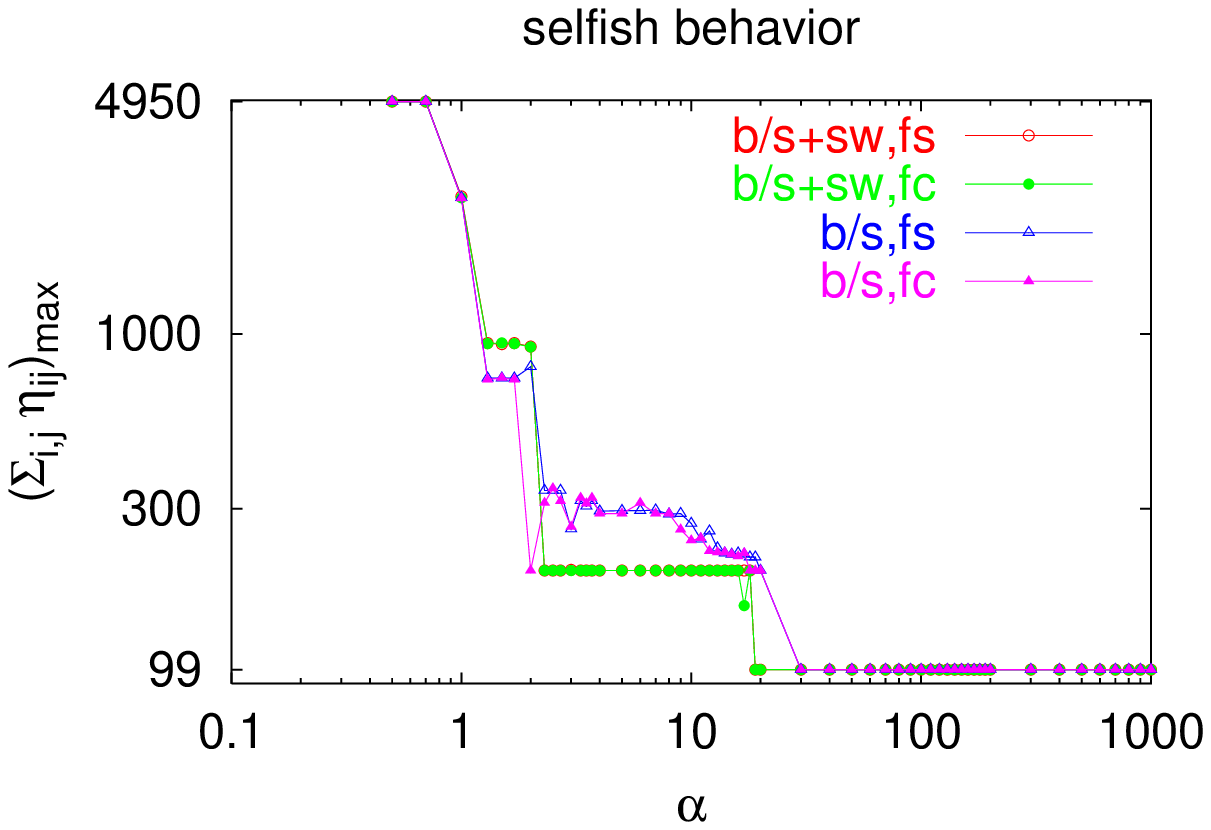}
\hfill
\includegraphics[width=.49\textwidth]{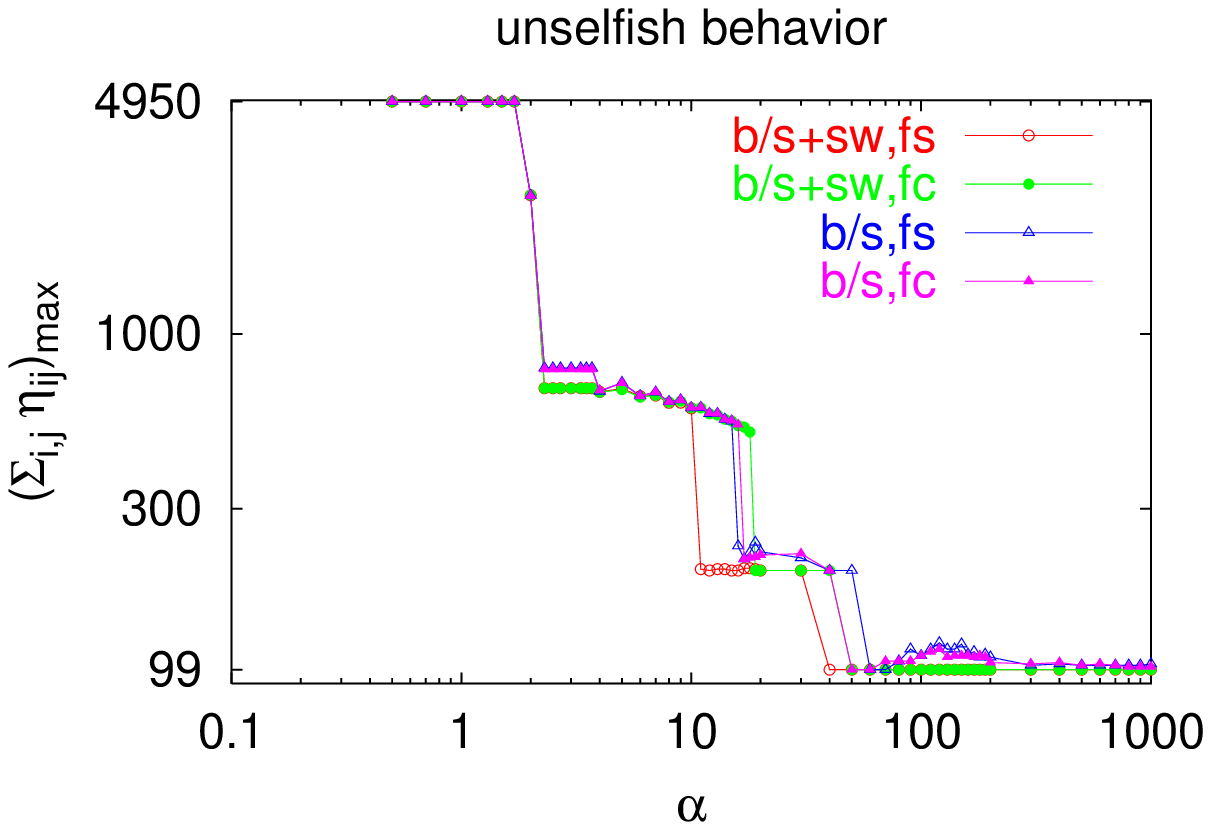}
}
\caption{Minimum (top), mean (middle), and maximum (bottom) number
of links in the final configurations
}
\label{fig:anzlinks}
\end{figure}

Next we have a look at the minimum, mean, and maximum number
of links in the final configurations, which are shown in
Fig.\ \ref{fig:anzlinks}. We get similar pictures here for the
various scenarios: for $\alpha<1$, we generally get completely
connected graphs with $N \times (N-1) / 2$ edges. As we always
use systems with $N=100$ nodes, this number is 4950. For large
values of $\alpha$, the number of edges is always given by
$N-1$, such that we always have trees here. In the intermediate
regime, there is a gradual decrease of the number of edges in
the system, which is nearly monotonous. For selfish agents
using both the buy/sell-move and the switch-move, we get a large
plateau for the maximum number of links at intermediate values of
$\alpha$. This maximum number is given by 196, i.e., by
$2 \times (N-3) +2$, the corresponding structure is the
graph with three central nodes which was shown in Fig.\
\ref{fig:examples}.
 
\begin{figure}\centering
\parbox{\textwidth}{
\includegraphics[width=.49\textwidth]{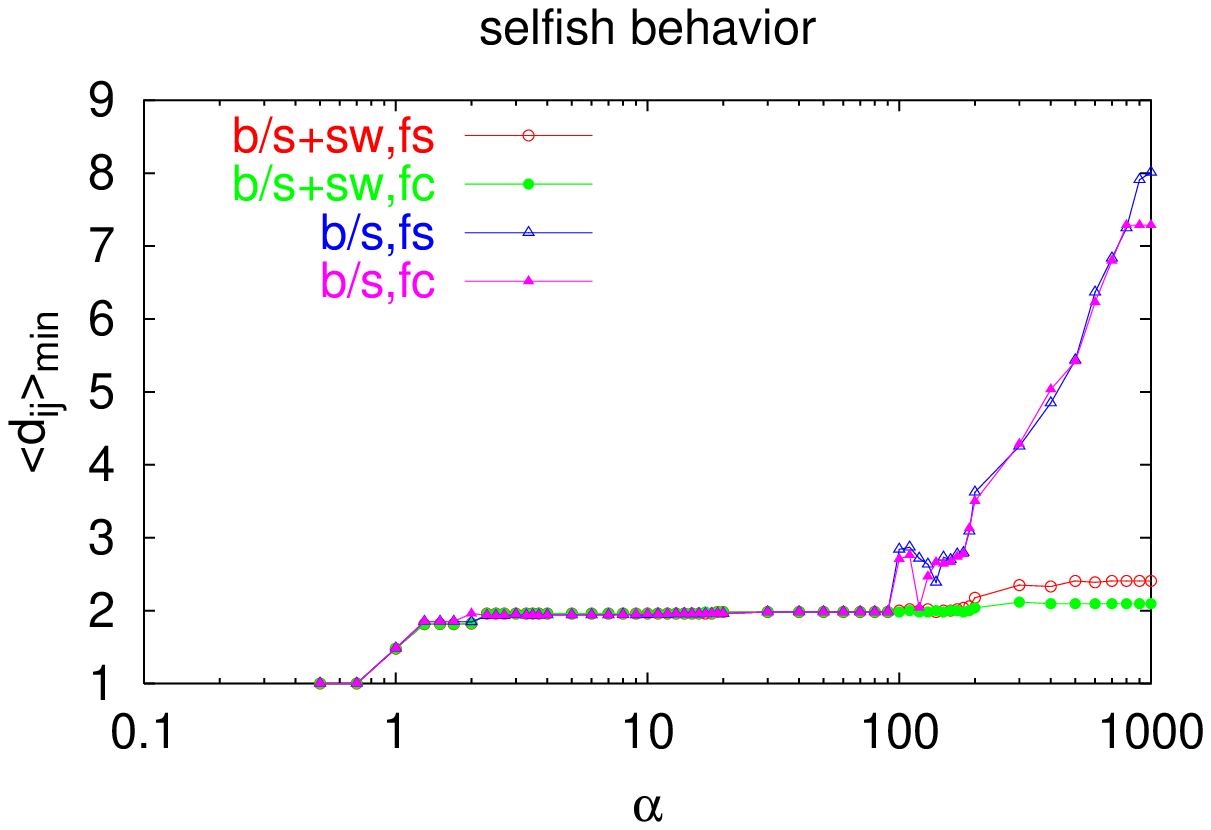}
\hfill
\includegraphics[width=.49\textwidth]{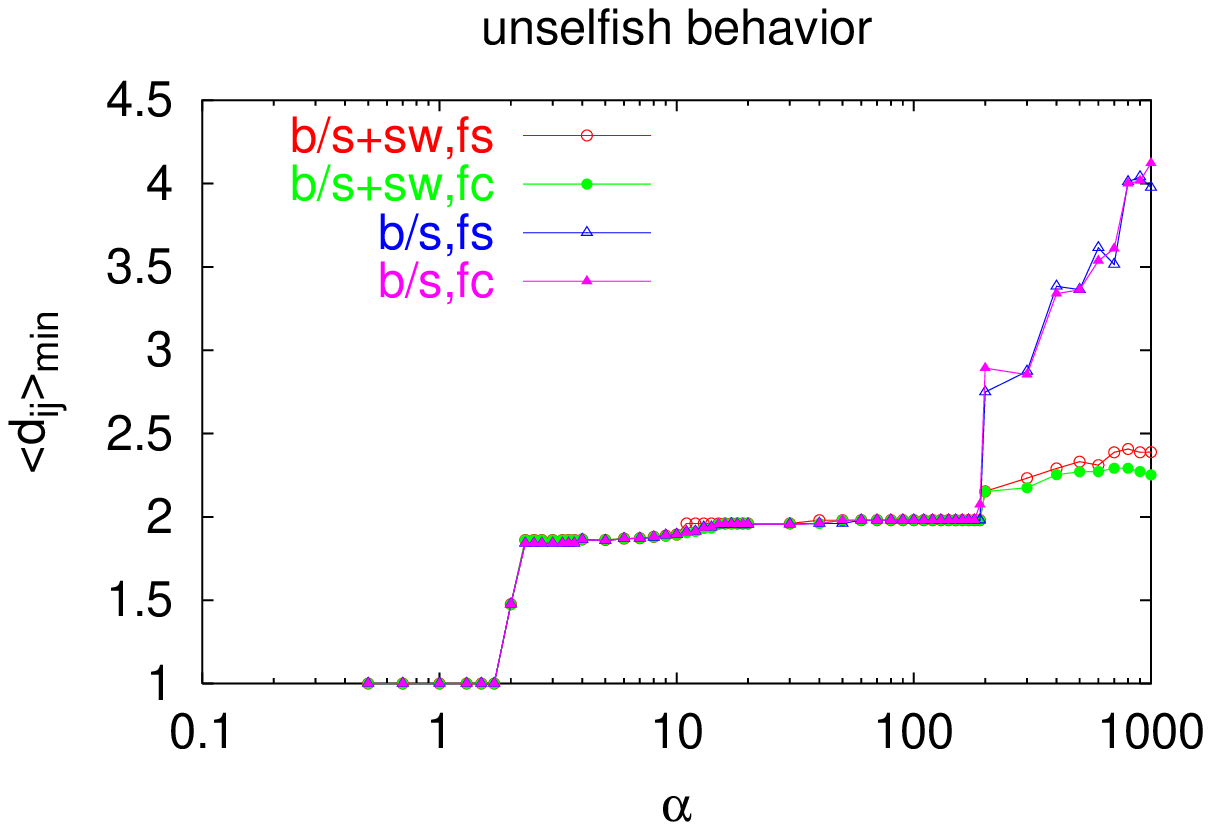}
}
\parbox{\textwidth}{
\includegraphics[width=.49\textwidth]{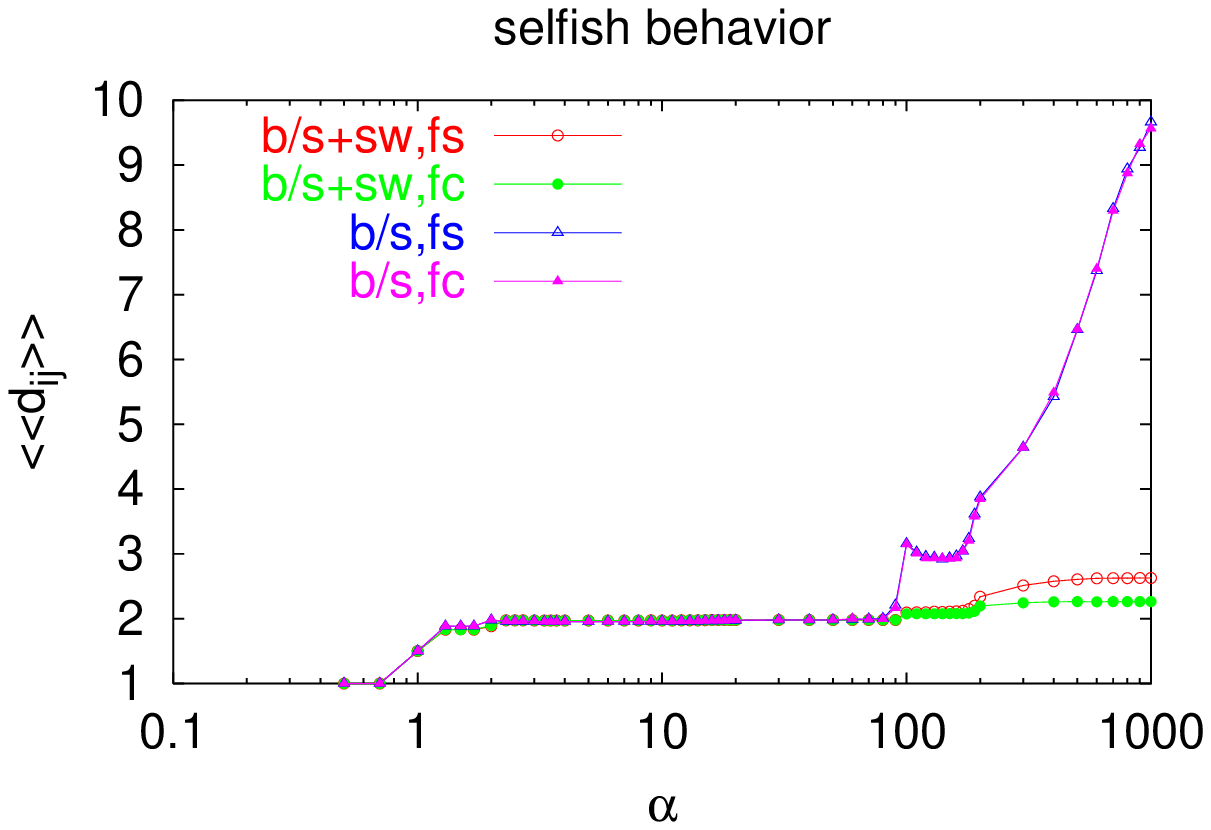}
\hfill
\includegraphics[width=.49\textwidth]{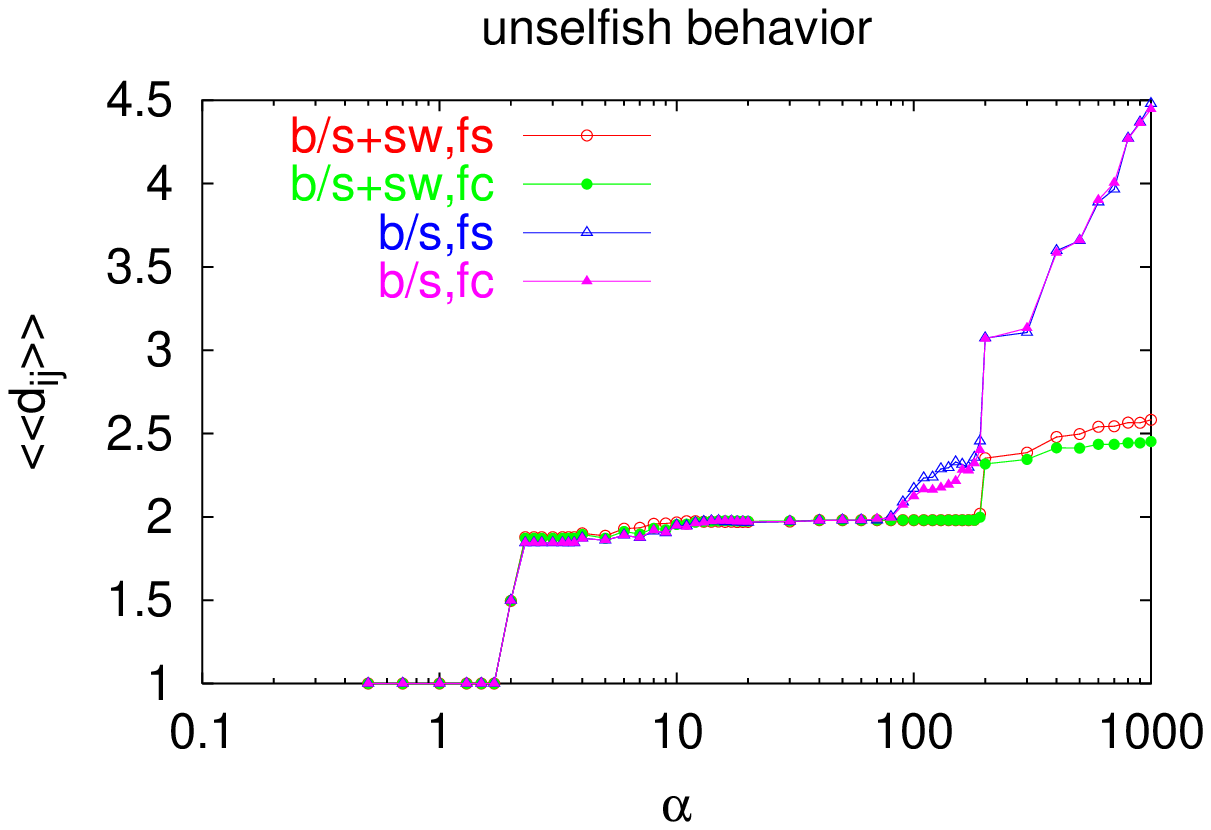}
}
\parbox{\textwidth}{
\includegraphics[width=.49\textwidth]{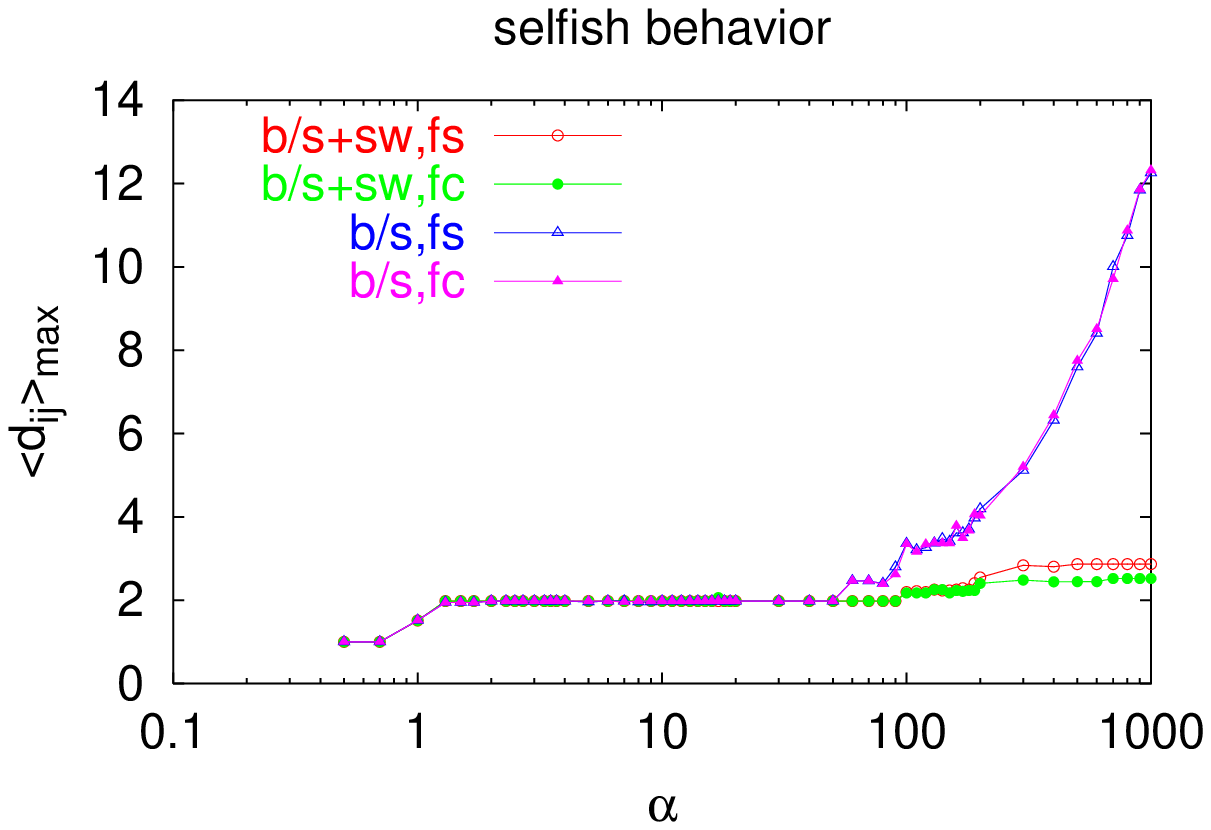}
\hfill
\includegraphics[width=.49\textwidth]{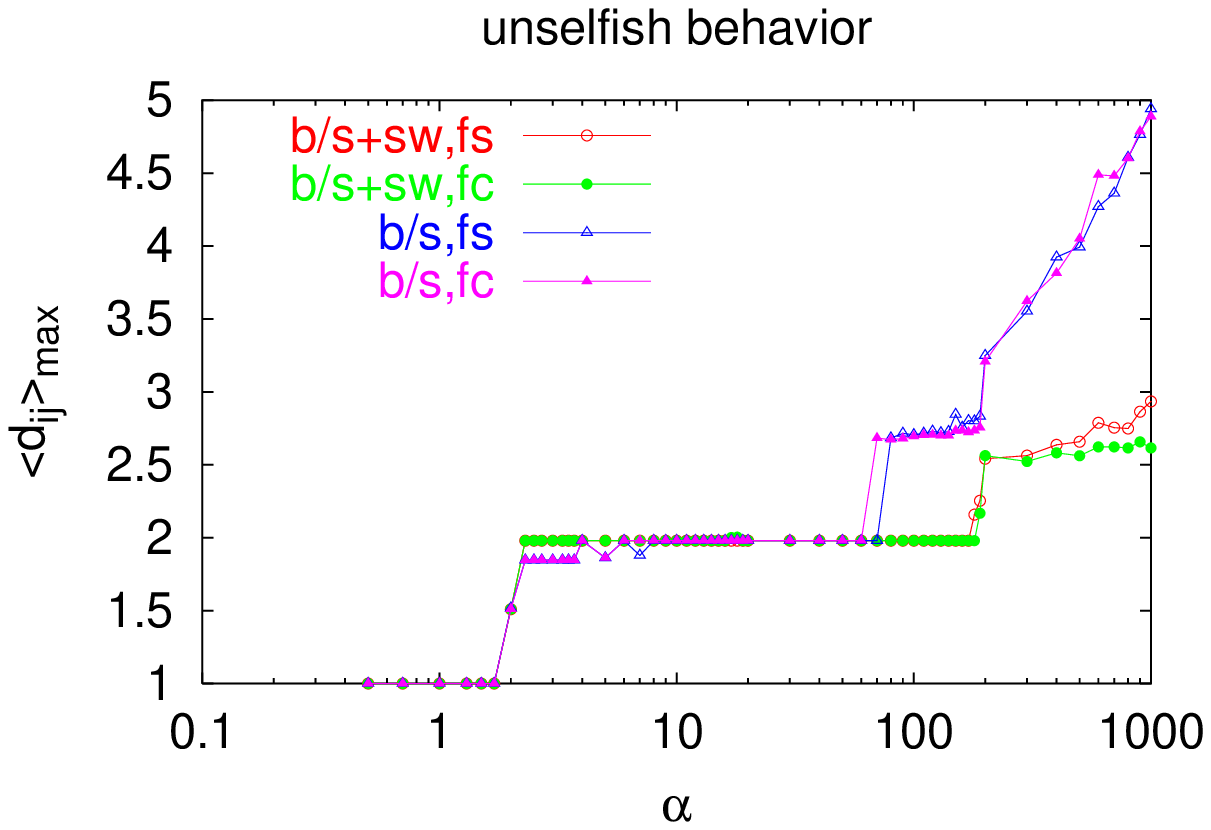}
}
\caption{Minimum (top), mean (middle), and maximum (bottom) average
distance in the final configurations
}
\label{fig:distances}
\end{figure}

Furthermore, we are interested in the average distance between
two different nodes in the network, which is measured in the
number of hops a message needs to get from one node to the other.
Looking at Fig.\ \ref{fig:distances}, we find three different
regimes depending on the value of $\alpha$: for small $\alpha$,
we always get the completely connected graph, such that the
average distance between two nodes is 1. Increasing
$\alpha$, the average distance increases to a value of $\approx 2$.
We get rather a plateau at this value for intermediate $\alpha$.
For large $\alpha$, the average distance increases again.
In this regime, we find a significant difference between the simulations
working both with the buy/sell-move and the switch-move or using
only the buy/sell-move: without the switch-move, the average distance
explodes with increasing $\alpha$, whereas the increase is
comparatively small for those simulations in which the switch-move
was implemented. If using both moves, selfish and unselfish behaviour
of the agents leads to roughly the same average distances.
But unselfish behaviour pays off if using only the buy/sell-move.

This result can be interpreted from two different points of view:
from the point of view of a moralist, one can say that unselfish
behaviour is superior to selfish behaviour and thus leads to overall
better results for all agents. Therefore, one should never allow
own egoisms to dictate the decisions one makes. On the other hand,
from the point of view of optimization, one has to state that
in the case of the unselfish agents, basically the cost function of the
overall system was considered when performing a move. Thus, every
move tried to minimize this overall cost function, whereas the
selfish agents worked only on their local part of the cost function.
Therefore, from the point of view of an optimizer, it is quite
clear that unselfish optimization has to lead to better or at least
equally good solutions than selfish optimization.

\begin{figure}\centering
\parbox{\textwidth}{
\includegraphics[width=.49\textwidth]{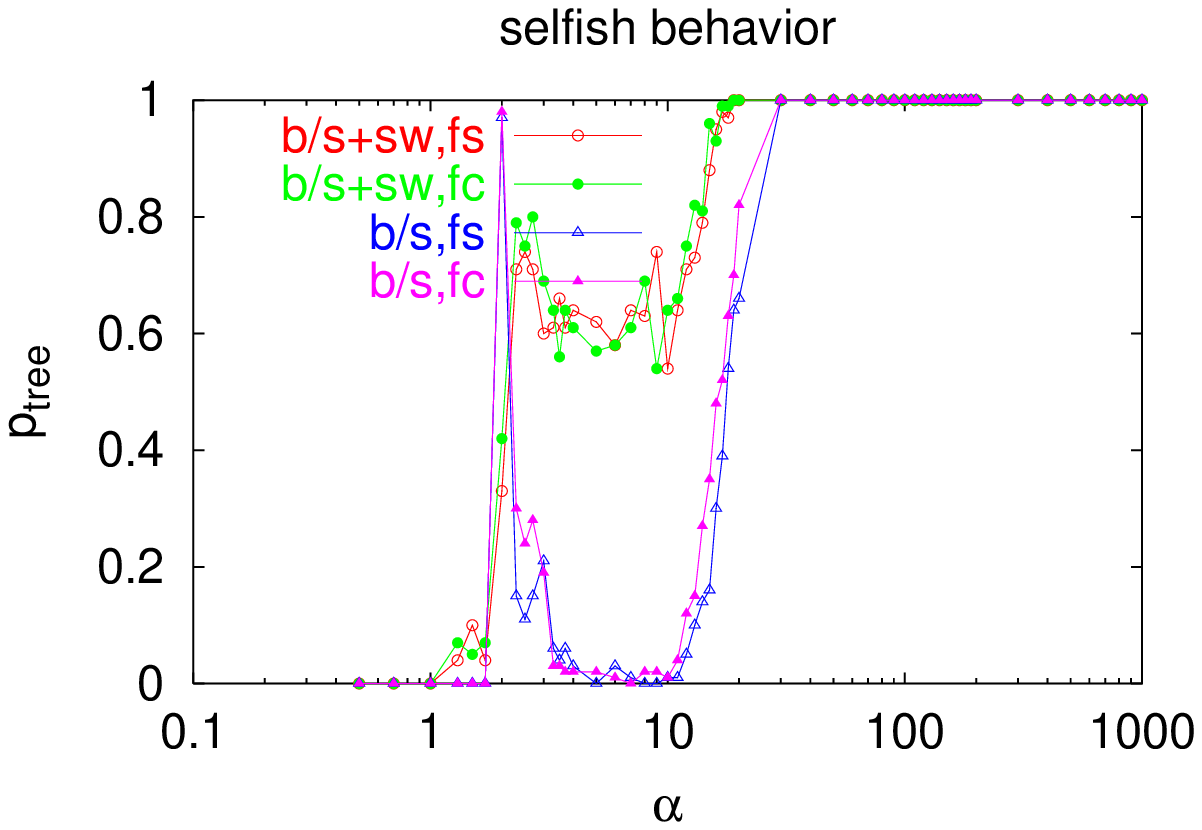}
\hfill
\includegraphics[width=.49\textwidth]{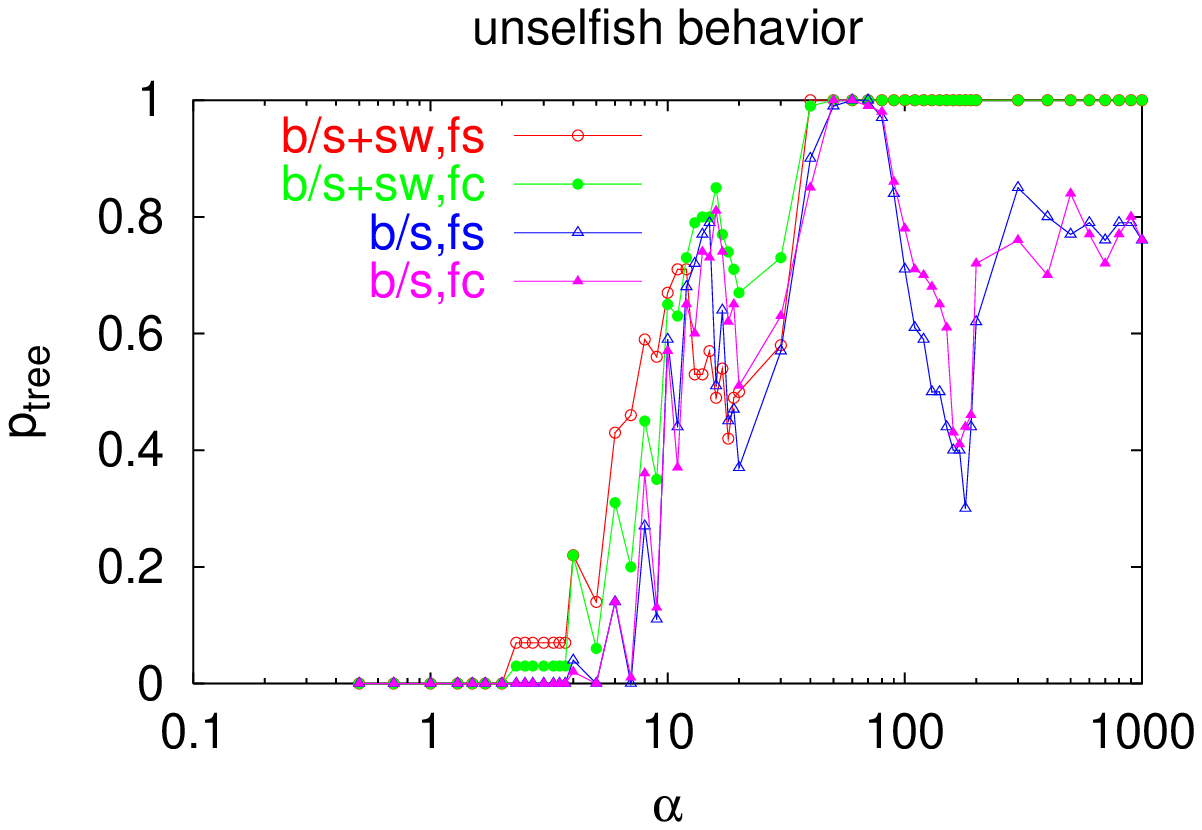}
}
\caption{Probability that the resulting configuration is a
tree}
\label{fig:tree}
\end{figure}

\begin{figure}\centering
\parbox{\textwidth}{
\includegraphics[width=.49\textwidth]{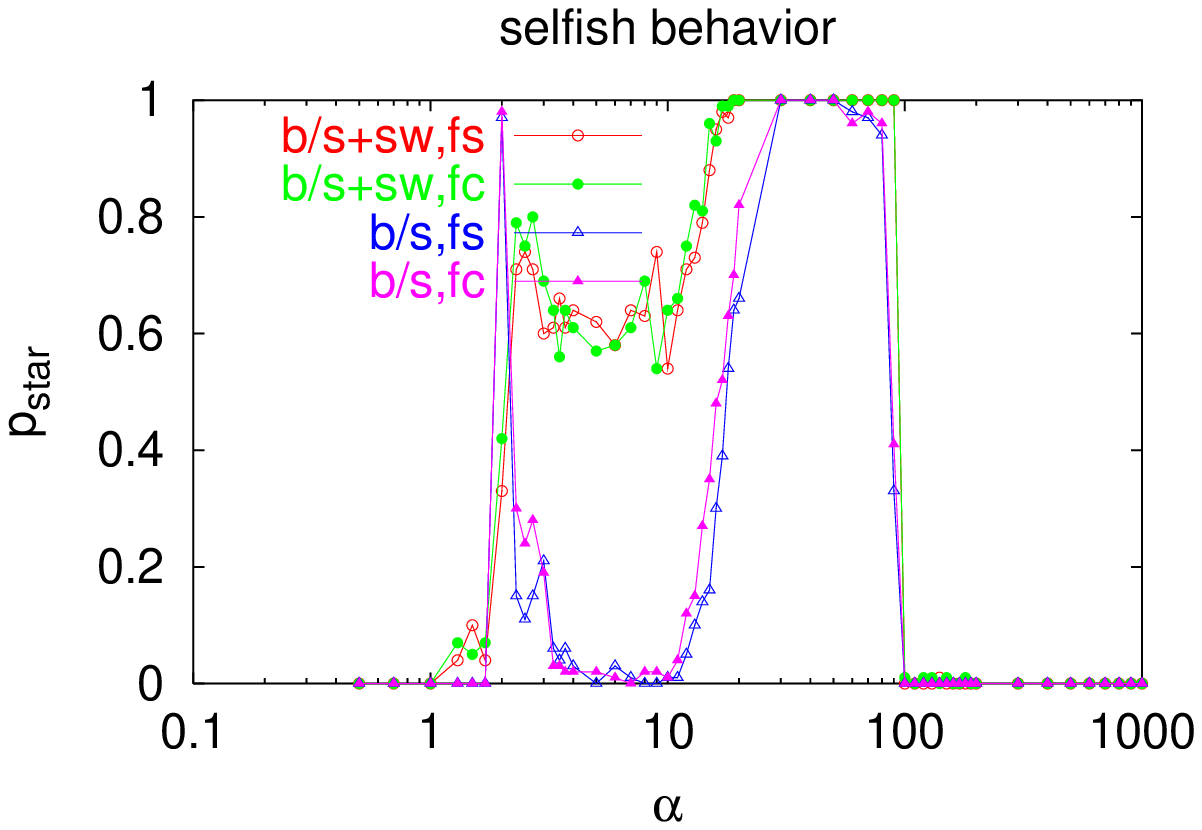}
\hfill
\includegraphics[width=.49\textwidth]{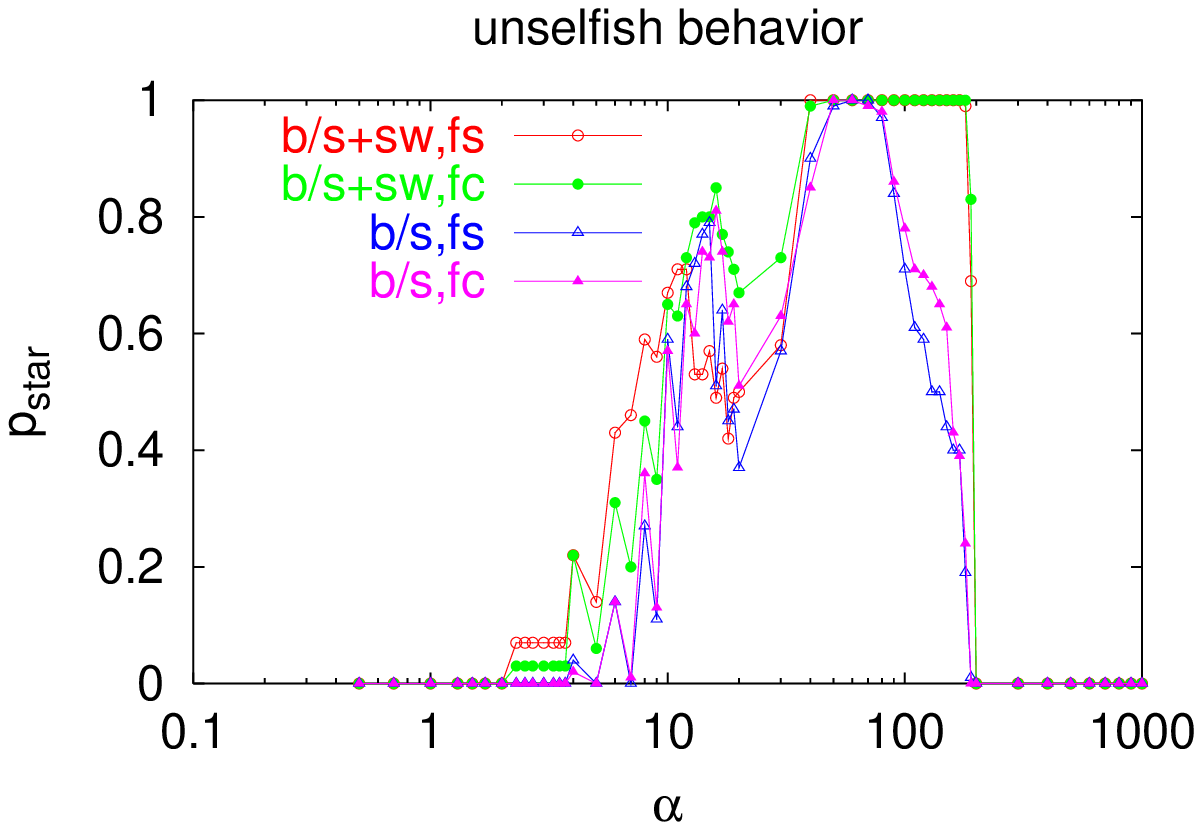}
}
\caption{Probability that the resulting configuration is a
star}
\label{fig:star}
\end{figure}

Finally, we have a look at the probabilities whether the
resulting configuration is a tree or a star, which are shown
in Figs.\ \ref{fig:tree} and \ref{fig:star}. For selfish agents,
the probability that the resulting configuration is a tree
is nearly 1 at $\alpha=2$ if only the buy/sell-move is used.
In this case, the probability drops to zero if increasing $\alpha$
and then increases again for $\alpha\ge 10$. For $\alpha\ge 30$,
it is again equal to 1.
If also using the switch-move, the probability increases from zero
to $\approx 0.75$ at $\alpha\approx 2.5$, then decreases and
stays at roughly a plateau of $\approx 0.6$ for $3 \le \alpha \le 10$,
and then increases to 1, reaching this value at $\alpha\approx 20$.
The corresponding picture for the probability that the resulting
configuration is a star is rather the same, except that the
probability breaks down from 1 to 0 at $\alpha\approx N$.

In the case of unselfish behaviour, we get a different picture
for these probabilities: the probability increases from 0 to
1 in the range $2 \le \alpha \le 50$ with a short breakdown
at $\alpha=20$. For large $\alpha$, we get different results:
if using only the buy/sell-move, the probability that the
resulting configuration is a tree breaks down when $\alpha$
approaches $N$ and finally fluctuates around $0.8$ for large
$\alpha$. If also working with the switch-move, the probability
stays 1. Again the picture is the same for the probability that
the final configuration is a star except that there the
probability breaks down for $\alpha>N$. This decrease is
performed gradually if only the buy/sell-move is used and
abruptly is both mobes are used.

Summarizing, if a simulation run ends in a tree for $\alpha<N$,
then it is always a star, whereas the probability for a star
vanishes for $\alpha>N$.

\section{Conclusion and Outlook}
In this publication, we have explored the structures generated when
multiple agents construct a computer network, within a highly simplified
model of the decisions that are made to achieve this.  We find that some
persistent and rather complex structures emerge in the intermediate regimes
of the parameters that characterize the network building process.  In
addition to the complete graph, and the star, the two idealized optimum
solutions (the first too expensive to be a realistic solution and the
second too fragile to be a reliable solution for a large scale network) we
find trees and multi-centred stars resulting from the process of network
formulation which we simulate.  The trees are more easily formed than a
star, and while loss of a single node separates the tree into a few parts
that can communicate only within each part, there is no central site which
controls all communication, as in the star.  The multi-centred star and
other approximate solutions that result from the simulations offer the germ
of a more reliable approach to network formation -- no pair of sites is
very far apart, and each pair can communicate over multiple distinct routes.

Of course, this network creation model could be criticized
in various ways for not being close to reality: for example,
the agents although trying to keep their costs small have
an infinite amount of money. Secondly, the cost for buying
a link is simply a constant and does not depend on the distance
between two nodes or on the bandwidth. Here it is also unrealistic
that a link in only owned by one of the two adjacent nodes and
that every node is free to send messages via this link.
Third, the bandwidth
problem within real networks is not considered at all in this
model. Instead, the model emphasizes only the number of hops
a message needs to get from the sender to the receiver,
which is often only a side-issue in real networks. Finally, the
configurations which emerge from this network are quite
unrealistic. Especially, the power law distribution in
the connection numbers of the nodes in real networks
is not reproduced in the model.

Our most interesting result for this model is the
comparison between the outcomes of selfish or
unselfish behaviours of the agents. From the point of
view of optimization, one would prefer to work with
unselfish agents as these basically consider the overall cost
function of the problem, which is to be minimized.
Contrarily, selfish agents only consider some local
part of the cost function.
However, working with selfish agents saves a lot of computing
time: as only the distances from one agent to all other
agents have to be evaluated for selfish agents, it
takes a computing time of the order ${\cal O}(N)$,
whereas all distances have to be evaluated for unselfish
agents, taking time of the order ${\cal O}(N^2)$.
Moreover, even if comparing the computing time in
number of steps or of moves, the simulations with
selfish agents often outperform those with unselfish
agents. If we define the quality of a network by the
average distance in hops between two arbitrary nodes,
which this network creation model attempts to optimize,
we find that we get equally good configurations for
small and intermediate values of $\alpha$ for both
types of behaviours. Only for large $\alpha$, simulations
with unselfish agents lead to better configurations.
Thus, we can summarize that simulations with selfish
agents are mostly superior to simulations with unselfish agents
in the way that they lead to equally good results in
much shorter computing times.

The model should serve as a starting point for future, more detailed
investigations, with more plausible models for the message traffic
required, or a more concrete description of the history and hierarchy by
which the actual Internet has been formed.

\end{document}